\documentclass[11pt]{article} 
\usepackage{geometry} 
\usepackage{bbm}
\usepackage{graphicx} 
\usepackage{color}
\usepackage{epstopdf} 
\usepackage{amsmath, amssymb} 
\usepackage{enumitem}[shortlabels] 
\usepackage{textcomp} 
\usepackage{setspace}
\usepackage{wrapfig}
\usepackage{float}
\usepackage{floatflt}
\usepackage{relsize}
\usepackage[parfill]{parskip} 
\usepackage{verbatim} 
\usepackage{bibentry} 
\usepackage{booktabs} 
\usepackage{bm}       
\usepackage{lineno}
\usepackage{authblk} 
\usepackage[margin=0pt,size=footnotesize, labelfont=bf, labelsep=colon]{caption}
\usepackage{sidecap}
\usepackage{algorithm}
\usepackage{algpseudocode}
\usepackage{algorithmicx}
\usepackage{colortbl}
\usepackage{xcolor}
\colorlet{tablerowcolor}{gray!10} 

\usepackage{multicol}

\usepackage{palatino}

\usepackage[numbers,super,sort&compress]{natbib}
\usepackage{hyperref}
\hypersetup{
allbordercolors = {white}
}

\makeatletter
\renewcommand\@biblabel[1]{#1.}
\makeatother

\newcommand{\bi}{\begin{itemize}}
\newcommand{\ei}{\end{itemize}}

\usepackage{empheq}
\usepackage[most]{tcolorbox}  

\newcommand{\ie}{\textit{i.e.}~}
\newcommand{\etal}{\textit{et al}}

\usepackage[normalem]{ulem}
\usepackage{url}
\DeclareUrlCommand\ULurl{%
  \renewcommand\UrlLeft{\uline\bgroup}%
  \renewcommand\UrlRight{\egroup}}

\title{Spatiodynamic Inference Using Vision-Based\\ Generative Modeling}
\author{Jun Won Park, Kangyu Zhao, Sanket Rane \thanks{Corresponding Author: sanket.rane@columbia.edu}\\
Irving Institute for Cancer Dynamics, Columbia University, New York, USA}
\date{}

\begin{document}
\maketitle

\begin{abstract}

Biological systems commonly exhibit complex spatiotemporal patterns whose underlying generative mechanisms pose a significant analytical challenge.
Traditional approaches to spatiodynamic inference rely on dimensionality reduction through summary statistics, which sacrifice complexity and interdependent structure intrinsic to these data in favor of parameter identifiability.
This imposes a fundamental constraint on reliably extracting mechanistic insights from spatiotemporal data, highlighting the need for analytical frameworks that preserve the full richness of these dynamical systems.
To address this, we developed a simulation‑based inference framework that employs vision transformer‑driven variational encoding to generate compact representations of the data, exploiting the inherent contextual dependencies.
These representations are subsequently integrated into a likelihood‑free Bayesian approach for parameter inference. The central idea is to construct a fine‑grained, structured mesh of latent representations from simulated dynamics through systematic exploration of the parameter space.
This encoded mesh of latent embeddings then serves as a reference map for retrieving parameter values that correspond to observed data.
By integrating generative modeling with Bayesian principles, our approach provides a unified inference framework to identify both spatial and temporal patterns that manifest in multivariate dynamical systems.
\end{abstract}

\clearpage 

\section*{Introduction}

Spatiotemporal patterns are ubiquitous in  biological systems whose mechanistic underpinnings present a significant analytical challenge.
Spatial dynamical data exhibit multi-scale dependencies among observed variables, encompassing both local and global relationships. 
Mathematical models are a natural language for describing such interconnectedness and for inferring causal relationships. 
However, the accuracy of their conclusions hinges on precisely identifying model parameters $(\theta)$ and quantifying the uncertainty in their estimation using the observed data $(y)$.

In Bayesian regimes, this involves determining the posterior distribution $\pi(\theta | y)$ given the prior $\pi(\theta)$ and the likelihood function $p(y | \theta)$.
In cases where traditional likelihood-based methods are impractical,  approximate Bayesian Computation (ABC) offers an appealing solution to estimating posterior using simulated data~\citep{Rubin_1984, Diggle_Gratton_1984, tavare1997inferring}.
ABC and its modern variants~\citep{Sisson_2007, marjoram2003markov, toni2009approximate, robert2008adaptivity} use a distance metric , $\rho (\cdot)$, to gauge the similarity between simulated and observed datasets and a threshold $\epsilon$ to accept posterior samples~(\textbf{eq.~\ref{eq:ABC}}), such that, as $\epsilon \rightarrow 0$, ABC posterior approaches the Bayesian posterior.
\begin{equation}
\pi(\theta | Y=y^{obs}) \approx \pi(\theta | \, \rho ( y^{sim}, y^{obs} ) \le \epsilon )
\label{eq:ABC}
\end{equation}

Inference on spatial data typically involves mixed models, consisting a regression model for covariates and a random-effects model, such as the Gaussian process, for spatial dependence~\citep{Cressie_2022, Kang_2011, Zhan_2024, Saha_2021, Datta_2016,Hamelijnck_2021}.
In many studies, specifically in epidemiology, summarized statistics are  leveraged to estimate model parameters using Bayesian methods~\citep{Brown_2018, Freitas_2021, Li_2023}.
Formulating meaningful summary statistics becomes increasingly difficult when working with high-dimensional spaces, frequently resulting in ill-posed inverse problems where parameter identifiability is compromised.
These factors critically impede reliable mechanistic inference on spatiotemporal dynamics, highlighting the need for more sophisticated approaches that preserve the full richness of these data.

To bridge this gap, we defined an integrative approach that combines a generative deep learning model with ABC to infer the underlying data-generating process and assess uncertainty in its parameter estimation.
We begin with a simple hypothesis that complex observed data contain underlying simpler patterns that are unobserved.
These hidden patterns can yield significant insights regarding the parameters that govern dynamical or mechanistic variations within the data. 
Our objective is to construct a structured mesh of latent representations of simulated data by systematically drawing samples from the prior.
This reference map will then be employed in a Bayesian framework to extract model parameters corresponding to the actual observations. 

We developed a variational inference-assisted ABC (viaABC) framework by integrating elements of Vision Transformer (ViT) and variational autoencoding (VAE) to construct the reference map of latent representations.
viaABC begins with a self-supervised training phase on synthetic data generated using parameter samples drawn from model-agnostic, weakly informative priors.
The trained encoder is integrated into a simulation-based inference framework to systematically compare the latent representations of observed data with those generated from simulations of the mechanistic model~(\textbf{eq.~\ref{eq:encode-abc}}) using an adaptive acceptance criterion~\cite{simola2021adaptive}.
Through this process, \mbox{viaABC} approximates the posterior distribution of the parameters that underlie the observed dynamics,
\begin{equation}
\pi(\theta | y=y^{obs}) \approx \pi(\theta | \, \rho ( \zeta(y^{sim}), \zeta(y^{obs}) ) \le \epsilon ), \\
\label{eq:encode-abc}
\end{equation}
where $\zeta(\cdot)$ represents variational encoding process that generates the latent variables.

\subsection*{Toy example: The spatial SIRS model}

\begin{figure}[h!]
    \centering
    \includegraphics[width=0.85\textwidth]{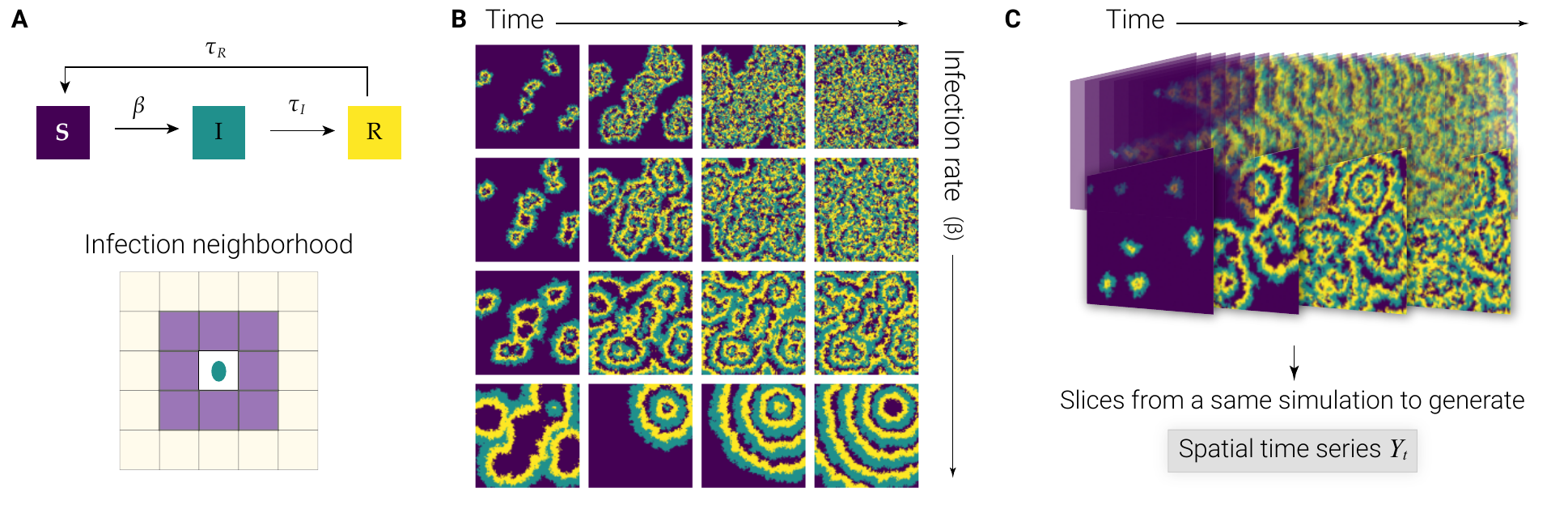}
    \vspace{3mm}
    \caption{\textbf{Spatial dynamics of the SIRS model}.
    \textbf{A.} Schematic of the state transitions. In addition, we depict the contact-grid for each individual (pixel), which serves as the infection neighborhood.
    \textbf{B.} An illustration of the varied patterns generated by the SIRS model. 
    This particular example depicts the output of SIRS model at times $t= 4, 8, 12, 16$ (left to right) by fixing $\tau_I = \tau_R = 1$ and varying $\beta = 0.4, 0.7, 1.0, 2.0$ (top to bottom). 
    Such diversity allows us to train the deep learning model on a wide assortment of spatiotemporal patterns, which we generate by sampling $\theta$ from a model agnostic prior distribution.
    \textbf{C.} Schematic of data generation and processing . We simulate the SIRS model using the time-step of $0.05$ and slice the generated video at every 20th step to construct the  spatial time series.}
    \label{fig:spatial_sir}
\end{figure}

Infectious disease spread is characterized by complex spatiotemporal patterns~\cite{Pei_2018, Shi_2022, Jalal_2024}, which
can be modeled through spatial adaptations of the susceptible-infected-recovered (SIR) model~\citep{Kermack_McKendrick_1927}.
To assess latent-ABC's ability to conduct mechanistic inference on spatial dynamical data, we employed a grid-structured contact network spatial SIR model as developed by van Ballegooijen \etal.~\citep{van_Ballegooijen_2004}.
This framework utilized a modification of the classical SIR, known as the SIRS model~\citep{Anderson_May_1991}, in which individuals occupy one of three states: susceptible (S), infected (I), or resistant (R)~(\textbf{Fig.~\ref{fig:spatial_sir}A}).

\vspace{2mm}

Within this model, infected individuals stochastically transmit the infection to neighboring susceptible individuals at the rate $\beta$.
The infection neighborhood is represented by eight adjacent individuals on a square lattice~(\textbf{Fig.~\ref{fig:spatial_sir}A}). 
The duration of the infectious period is fixed at $\tau_{I}$, following which the infected individuals transition to the resistant state. Subsequently, resistant individuals revert to the susceptible state after a fixed time period $\tau_{R}$. 
We adhered to the methodology outlined in van Ballegooijen \etal.~\citep{van_Ballegooijen_2004},  scale $\tau_{R}$ to unity and sought to infer $\theta = (\beta, \tau_{I})$, using our framework.
The SIRS model produces well-defined spatiotemporal patterns---from scattered clusters to infection waves---that are clearly delineated by its parameters~(\textbf{Fig.~\ref{fig:spatial_sir}B}) and, as such, provides an excellent test case for evaluating viaABC’s performance. 
We generated a synthetic dataset by fixing $\theta = (1.0, 1.0)$ as the ground truth.
We sampled 15 slices of images from this dataset to create a test dataset~(schematic shown in \textbf{Fig. \ref{fig:spatial_sir}B}), which we refer to as the `observed data' ($y^{obs}$).
We generated the training data for VAE similarly (see methods for details).

\subsection*{Cosine similarity of the latent variables as an acceptance criterion}
\begin{figure}[h!]
    \begin{center}
    \includegraphics[width=\textwidth]{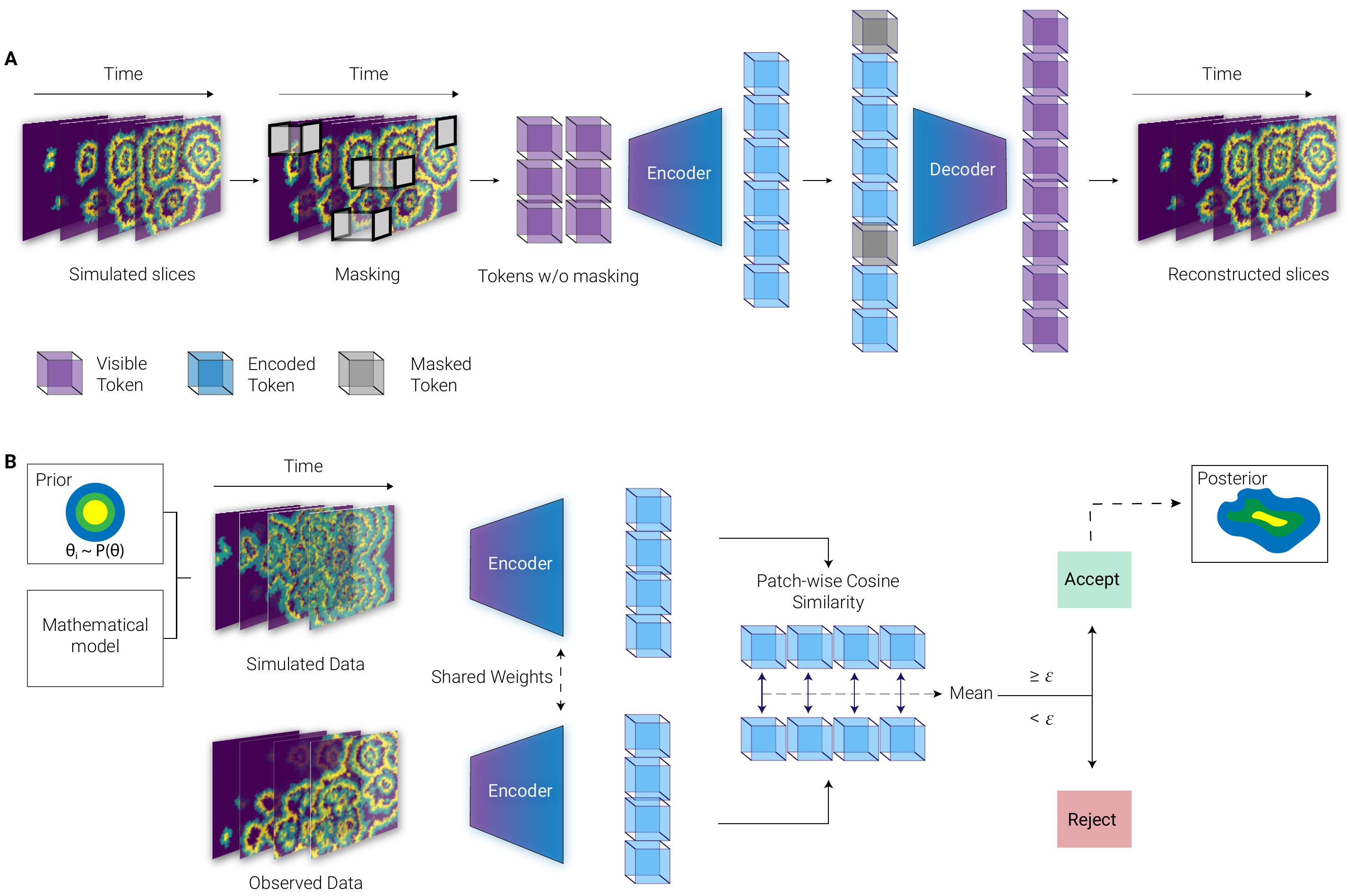}
    \end{center}
    \vspace{2mm}
    \caption{\textbf{Schematic of the viaABC framework}.
    We depict viaABC operation for spatial dynamical data relevant to the SIRS model.
    \textbf{A.~Training:}
    Each training sample consists of a sequence of simulated images or data slices. 
    Each slice is divided into multiple non-overlapping regions called patches.
    We group patches along the time dimension to form cubes, which are then flattened into vectorized representation called tokens.
    Within each training sample, a random subset of tokens is masked.
    VAE is trained to reconstruct the original data using tokens as the input, while minimizing the KL divergence loss in the latent space.
    Encoder is applied to only visible tokens (purple).
    Encoded tokens (blue) are probabilistically sampled and along with the mask tokens (gray) are then processed by a decoder.
  \textbf{B.~Inference:} Trained encoder is applied to the unmasked data, both simulated and observed, to infer model parameters using sequential Monte Carlo and patch-wise cosine similarity as an acceptance criterion.
  We exclude the decoder and sampling process from the inference procedure.
  Principles of training and inference remain same across all data types and architectures described in this study.
   }
   \label{fig:schematic}
\end{figure}

We postulate that contextual relationship between data points and patterns hidden within the data generating processes could be realized by projecting simulations and observed data to latent variables. 
Such projections could precisely inform on model parameters and may also help in model validation and selection procedures.
To facilitate this we propose to use closeness between latent variables of the observed and simulated data, which we evaluate using patch-wise cosine similarity, as an acceptance criterion~(\textbf{Fig.~\ref{fig:schematic}}).

We customized a VAE model by adopting the transformer-based masked autoencoder~\citep{feichtenhofer2022masked} and imposing a probability distribution in the latent space via Kullback–Leibler (KL) divergence. 
The resultant loss function is,
\begin{equation}
L = L_{\text{recon}} + \lambda L_{KL},
\label{eq:loss}
\end{equation}
where $L_{\text{recon}}$ is the reconstruction loss, $L_{KL}$ denotes the KL divergence loss. 
The hyperparameter $\lambda$ mediates the trade-off between $L_{\text{recon}}$ and $L_{KL}$~\citep{higgins2017beta}. 
Our approach considers the loss in reconstruction for both unmasked and masked patches, deviating from conventional masked modeling approaches that only rely on masked patches.~\citep{devlin2019bert, he2022masked, tong2022videomae}.
In addition, we only use the encoded tokens to calculate the KL divergence loss~(\textbf{Fig.~\ref{fig:schematic}A}).

To generate the training data we draw \( N \)  parameter sets from a joint prior distribution using Latin hypercube sampling (LHS).
We chose LHS as it provides a thorough exploration of the parameter space resulting in a diverse range of parameter configurations~\citep{mckay2000comparison}.
Further, we use wide uniform priors for the training procedure to mitigate the risk of overfitting to specific modes in the parameter space. 
Given a mathematical model or data-generating process \( f \), this procedure yields $N$ pairs of $\{ (\theta_i, f(\theta_i)) \}_{i=1}^{N}$.
Masked VAE is trained on $\{ f(\theta_i) \}_{i=1}^{N}$ to reconstruct the input data while minimizing the total loss $L$~(\textbf{Fig.~\ref{fig:schematic}A}).

We define ABC posterior following the notations of Marin et al.\cite{marin2012approximate} and Simola et al.\cite{simola2021adaptive} as,
$$
\pi_{\epsilon}(\theta | y^{obs}) = \int \bigg[\frac{p(y^{sim} | \theta)\pi(\theta)\mathbbm{1}_{A_{\epsilon, y^{obs}}}(y^{sim})}{\int_{A_{\epsilon, y^{obs}} \times \Theta } p(y | \theta)\pi(\theta)dy^{sim}d\theta} \bigg]dy^{sim},
$$
where $\mathbbm{1}_{A_{\epsilon, y^{obs}}}(\cdot)$ is the indicator function for the set $A_{\epsilon, y^{obs}} = \{y^{sim} | \rho (s(y^{obs}), s(y^{sim})) \leq \epsilon\}$, \\ $\rho(\cdot,\cdot)$ is the distance function and $s(\cdot)$ is the summary statistic function. 

In viaABC, we modify the set notation such that,
$$A_{\epsilon, y^{obs}} = \{y^{sim} | \rho (z^{obs}, z^{sim}) \leq \epsilon\},$$
where $z^{obs}$ and $z^{sim}$ are latent embeddings of the observed and simulated data outputted by the combination of the encoder and re-parameterization layers. 
The re-parameterization layer imposes a standard Gaussian distribution in the latent space through,
\begin{equation}
Z = \mu + \sigma \epsilon, \quad \epsilon \sim \mathcal{N}(0, I),
\label{eq:reparameterization}
\end{equation}
which then outputs mean $\mu_i$  and standard deviation $\sigma_i$ vectors for each token $i$.
Both the decoder and sampling process are excluded from the inference procedure and only the vector of means $\mu$ is considered~(\textbf{Fig.~\ref{fig:schematic}B}) such that the latent representation $Z = \mu$~(\textbf{eq.~\ref{eq:reparameterization}}).
We use patch-wise cosine distance metric to compute the distance between latent embeddings~(\textbf{Algorithm~\ref{alg:viaABC} Line 6}),  (\textbf{Fig.~\ref{fig:schematic}B}).
For each pair of corresponding patches in $z^{obs}$ and $z^{sim}$, we compute cosine similarity and aggregate across all $n$ patches such that
\begin{equation}
\rho (z^{sim}, z^{obs}) = 1 - \sum_{i=1}^{n} \frac{z_i^{obs} \cdot z_i^{sim}}{||z_i^{obs}||_2||z_i^{sim}||_2}
\label{eq:cosine-metric}
\end{equation}
where $Z \in \mathbb{R}^{n \times E}$, where $E$ is the embedding dimension of the decoder.
Sampled parameters are accepted based on the patch-wise cosine distance metric~(\textbf{eq.~\ref{eq:cosine-metric}}).

\subsection*{Parameter inference from learned representations of spatiotemporal input}
\begin{figure}[H]
    \centering
    \includegraphics[width=0.99\linewidth]{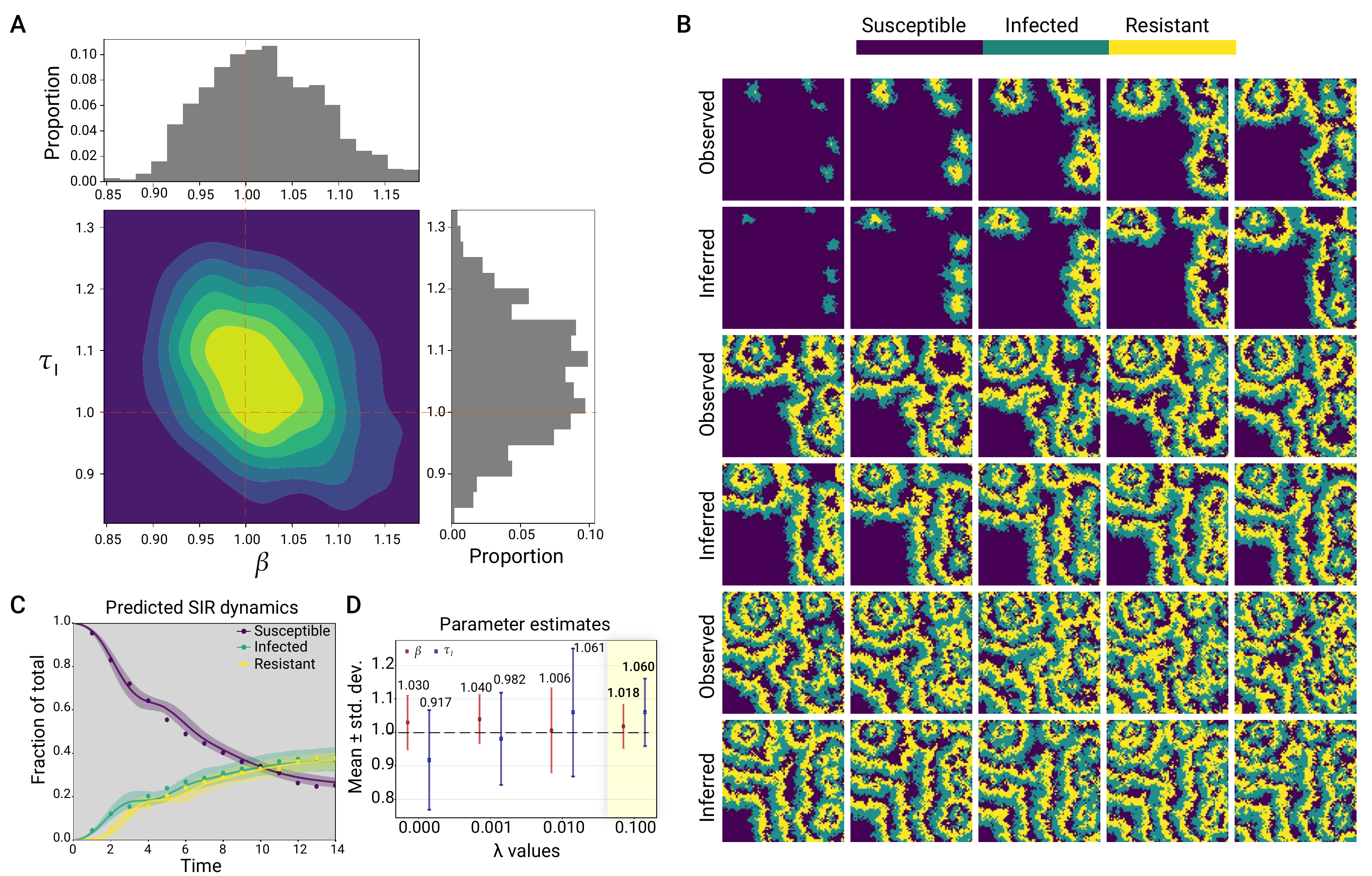}
    \caption{\textbf{Posterior inference on spatial SIRS dynamics using viaABC}.
\textbf{A.} Kernel density estimate of the final particles for the parameters $\beta$ and $\tau_I$.
\textbf{B.} Comparison of spatiotemporal dynamics of the observed data and predictions generated using the mean parameter estimates. Predictions were initialized within the similar region on the spatial grid as that of the observed data (see Methods for details).
\textbf{C.} SIRS dynamics aggregated across spatial dimensions. We average the proportions of S, I, and R at each time point across 10 simulations generated from each posterior draw. Dots represent observed data.
Lines denote the mean of predictions, and envelopes are 95\% credible intervals.
\textbf{D.} Mean estimates and corresponding standard deviations, calculated from five independent experiments for each $\lambda$ value.
Results from ViT-VAE trained with $\lambda = 0.1$ are highlighted in yellow. Dashed line represent the ground truth.
Note that $\lambda = 0$ corresponds to the standard MAE.
  }
    \label{fig:inference}
\end{figure}
We employed viaABC to estimate the parameters ($\theta = (\beta, \tau_I)$) of the spatial SIRS model using synthetically generated observed data~(\textbf{Fig.~\ref{fig:spatial_sir}C}).
We sampled $\beta$ and $\tau_I$ from a bivariate $\text{Unif}(0, 4.2)$ using LHS and generated 50,000 simulations for training and an additional 10,000 simulations for validation. 
Each simulation contains 15 time points, with each time point representing a tensor of dimension $3 \times 80 \times 80$.
Here $80 \times 80$ denotes the spatial grid and the first dimension encodes the state of each individual. 
The three channels reflect distinct states---susceptible, infected, or resistant---encoded in a binary format; for instance, each individual is denoted by a 1 in the second channel if infected, otherwise 0.
As the datatype only contained discrete labels, we formulated the reconstruction loss as cross-entropy loss.

The VAE architecture comprised a ViT based encoder with 6 layers and an embedding dimension of 128, and a  smaller ViT based decoder with 4 layers and an embedding dimension of 64.
The input data of size $3 \times 15 \times 80 \times 80$ is divided into non-overlapping spatio-temporal cubes of size $3 \times 3 \times 10 \times 10$, yielding 500 cubes.
Each cube is flattened to form a token, 15\% of which are randomly masked during training (as depicted in \textbf{Fig.~\ref{fig:schematic}B}). We set $\lambda=0.1$ in \textbf{eq.~\ref{eq:loss}} to calculate the loss and train the VAE.
We used the trained encoder and same prior as before to perform the posterior inference.
viaABC iteratively and adaptively accepted 1,000 independent particles until the stopping rule of $q_t = 0.99$ or $t \geq 20$ was satisfied~(as described in \textbf{Algorithm~\ref{alg:viaABC}}).

The posterior inferred through viaABC contained the ground truth, and the estimated mean $\hat{\theta} \approx (1.04, 0.98)$ nearly coincided with $\theta = (1, 1)$~(\textbf{Fig.~\ref{fig:inference}A} and \textbf{D}).
Consistent with this, the predicted spatial SIR dynamics, generated using the mean of the posterior, closely mirrored the observed data~(\textbf{Fig.~\ref{fig:inference}B}).
We further show that the summarized SIRS dynamics generated using posterior draws nicely captured the summary statistics of the observed data~(\textbf{Fig.~\ref{fig:inference}C}).
Lastly, we explored the effects of varying $\lambda$ on accuracy of the posterior inference.
We present the aggregated results across five independent experiments in (\textbf{Fig.~\ref{fig:inference}B}).
For all configurations, the 95\% highest density interval (HDI) from the posterior inference included the ground truth, suggesting that viABC is rather weakly sensitive to fluctuations in the regularization parameter $\lambda$.
Notably, all VAE configurations improved on the results from the autoencoder version (\textbf{Fig.~\ref{fig:inference}D}) and, on average, converged more efficiently, indicating a more structured latent space.

\subsection*{Inference on noisy dynamical input}

Next, we assessed and compared the efficacy of the viaABC algorithm in a non-spatial setting. 
We used the Lotka-Volterra model~\citep{lotka1925elements}\citep{volterra1928variations}, a classic system in ecology, which deterministically describes the population dynamics and  interactions between a prey $(x)$ and predator $(y)$ species.
This system is expressed as a pair of nonlinear differential equations,
\begin{equation}
\frac{dx}{dt} = a x - cxy, \qquad
\frac{dy}{dt} = b xy - dy,
\label{eq:LV}
\end{equation}
where $a$ and  $b$ denote rates of population growth of the prey and their per capita loss due to predation.
Parameters $c$ and $d$ represent the efficiency with which consumed prey are converted into predator offspring and loss of the predators, respectively.
The Lotka-Volterra model is a well-studied system, and its parameters have been estimated using various methods, including ABC~\citep{toni2009approximate, prangle2017adapting}, ABC with distributional Random Forest (ABC-DRF)~\citep{dinh2024approximate}, and ABC with deep learning (ABC-DL)~\citep{Jiang2017, akesson2021convolutional, Baragatti2024}.
We adhered to the methodology outlined in Toni~\citep{toni2009approximate} and Dinh~\citep{dinh2024approximate} \etal., fixed parameters $c=d=1$  and sought to infer \( \theta = (a, b)\) using the prior  \( a, b \sim \text{Uniform}(0, 10)\).

\begin{figure}[h!]
    \centering
    \includegraphics[width=0.9\linewidth]{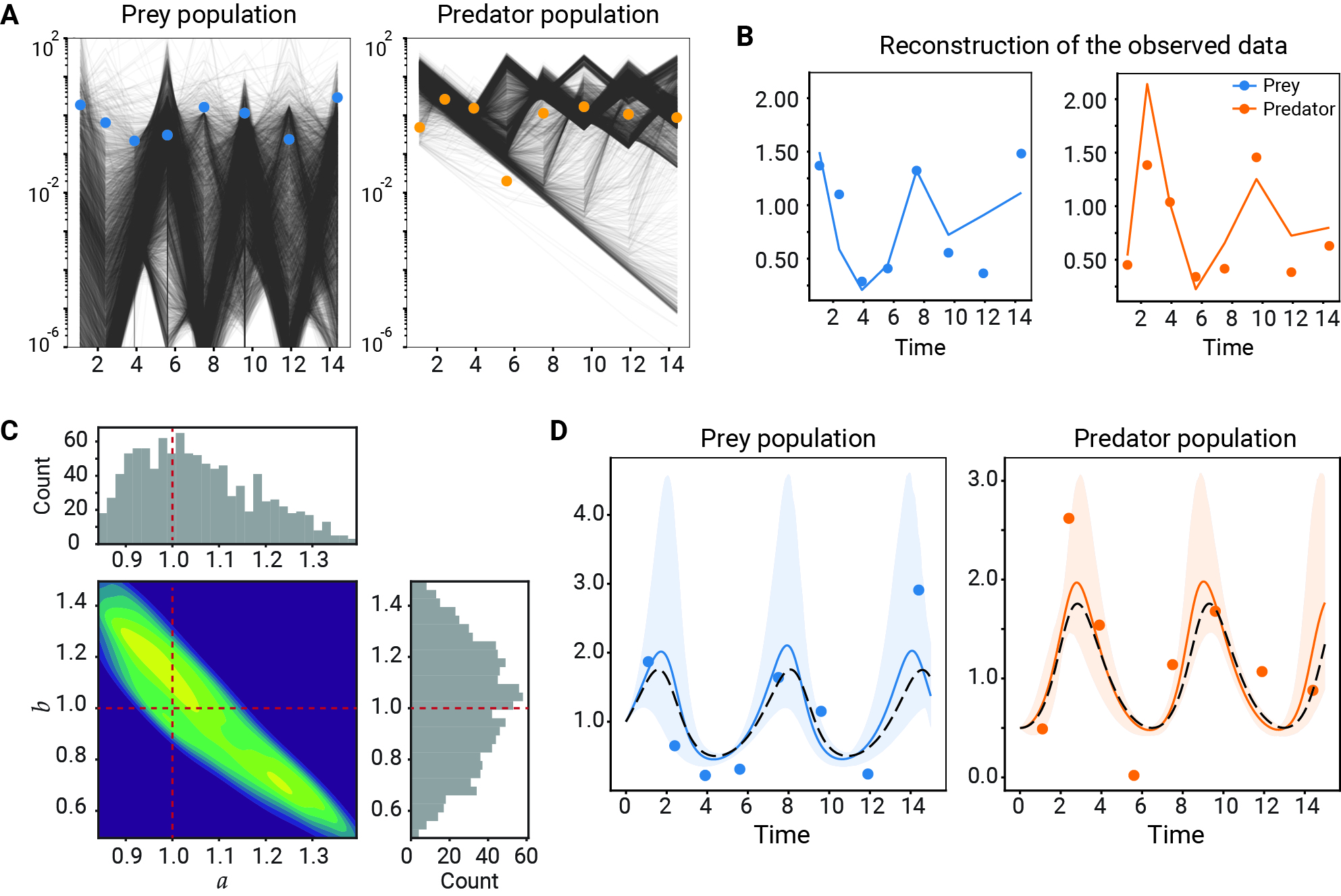}
\caption{\textbf{Learning and inferring predator-prey dynamics}. 
\textbf{A.} The simulated dynamics of the training data, denoted as lines, and the observed data, denoted as dots.
Y-axis is on the log scale to fully visualize the wide range of predator-prey dynamics in the simulated data.
\textbf{B.} ViT-VAE's reconstructions (lines) of the noisy observed data (dots).
\textbf{C.} The kernel density estimated 2D heatmap of the final posteriors along with their histograms.
The ground truth is denoted as dotted red lines.
The estimated 95\% HDI for $a, b$ are (0.951, 1.188) and (0.768, 1.352), respectively.
\textbf{D.} The 95\% credible intervals of the predictions of prey, predator dynamics with medians as solid lines.
Black dashed lines denote predator-prey dynamics generated using the ground truth without the Gaussian noise.}
    \label{fig:lv}
\end{figure}

We generated $N=50,000$ training simulations and $20,000$ validation samples of prey, predator dynamics~(shown as lines in~\textbf{Fig.~\ref{fig:lv}A}), by sampling $\theta$ from the joint prior distribution using LHS.
We selected eight time points between $t=0$ and $15$ to represent our training data $s^{\mathrm{train}} = \{\{x^i_1, y^i_1, \dots, x^i_8, y^i_8\}\}_{i=1}^{N}$.
The input data of size $2 \times 8$ is divided into 8 non-overlapping patches of size $2\times 1$.
We trained the ViT-VAE using a masking ratio of 15\%. The comparison between reconstructions of masked and unmasked training data shows that our ViT-VAE  effectively imputes missing observations using the patterns within the training data~(\textbf{Fig.~\ref{fig:masking}}).

To construct the observed data, we evaluated $x, y$ using eq.~\ref{eq:LV} at the same eight time points as the training samples and by setting ($\theta = (1, 1)$, see methods for details).
Gaussian noise \(\epsilon_t \sim \mathcal{N}(0, 0.5^2)\) was added to this time series to represent experimental variation typically associated with biological samples.
This forms our noisy dynamical input $s^{\mathrm{obs}} = \{x^o_1, y^o_1, \dots, x^o_8, y^o_8\}$~(shown as dots in~\textbf{Fig.~\ref{fig:lv}A}), which we used as the observed data for simulation-based inference.
In~\textbf{Fig.~\ref{fig:lv}B}, we show the reconstruction of the $s^{\mathrm{obs}}$~(original data as dots, reconstructed as lines). 

To generate simulated data for parameter inference, we sampled $\theta$ from the bivariate uniform prior to simulate time series $s^{\mathrm{sim}} = \{x^s_1, y^s_1, \dots, x^s_8, y^s_8\}$.
Both, observed simulated data are then encoded into their respective latent variables and were used to infer posterior of $\theta$ using viaABC. 
Specifically, 
We accepted 1,000 particles based on the cosine similarity of $z^{\mathrm{sim}}$ to $z^{\mathrm{obs}}$.
This procedure is repeated until the stopping rule described in the~\textbf{Algorithm~\ref{alg:viaABC}} is satisfied.

We observed that the ground truth $\theta = (1, 1)$ lied within the high density region of the posterior of the accepted particles, very close to the estimated $\hat{\theta} \approx (1.06, 1.06)$~(\textbf{Fig.~\ref{fig:lv}C}).
We further show that the 95\% credible intervals of the predictions of prey, predator dynamics contains the data generated using the ground truth and the median of the interval closely overlaps with the observed data~\textbf{(Fig.~\ref{fig:lv}D}).
Lastly, we benchmarked our approach against other Monte Carlo-based approaches and found that viaABC delivered results comparable to well-known Bayesian algorithms~(\textbf{Table~\ref{tab:lv_benchmark}}) and produced lowest uncertainty in parameter estimation.

\begin{table}[h!]
\centering
\begin{tabular}{l cccc}
\toprule
Statistics      & viaABC & MCMC & ABC-SMC-DRF  & ABC-SMC \\
\midrule
$\mathbb{E}$(a) & \textbf{1.064} & 1.087           & 1.121  & 1.291          \\
Var(a)          & \textbf{0.005} & 0.006           & 0.033  & 0.110          \\
$\mathbb{E}$(b) & 1.057          & 1.100           & 0.970  & \textbf{1.026} \\
Var(b)          & \textbf{0.029} & 0.034           & 0.031  & 0.104          \\
\bottomrule
\end{tabular}
\caption{\textbf{Means and variances of marginal posterior distributions of the parameters of the Lotka-Volterra model.} 
We compared the estimates obtained using viaABC with Markov Chain Monte Carlo (MCMC) sampler (produced using \textit{stan} programming language and same priors, results shown in~(\textbf{Fig.~\ref{fig:stan_fit}}), ABC-SMC and ABC-SMC-DRF (results directly taken from Dinh \etal~\citep{dinh2024approximate}).
Expectations closest to the ground truth and lowest variances are highlighted in bold.}
\label{tab:lv_benchmark}
\end{table}

\section*{Discussion}
In this study we introduced viaABC, a simulation-based inference framework that uses contextual dependencies inherent in the data to estimate parameters of multivariate dynamical systems.
The framework employs a vision transformer-based masked variational autoencoder to create ordered latent representations of model simulations, where simulations with similar dynamics are positioned close together in the embedding space.
The key insight is that by systematically exploring the parameter space, viaABC builds a comprehensive reference library of simulated dynamics that helps to identify which parameters most likely generated the patterns we see in the observed data.
We demonstrated efficacy of this approach using the spatial stochastic SIRS system, which exhibits complex dependencies between spatial dimensions and across time, where viaABC accurately retrieved model parameters and quantified the uncertainty in their estimation.

Variational encoding confers viaABC the ability to accommodate datasets with diverse structures and dimensions, as VAE can be employed to compress or expand data dimensions as needed for parameter inference.
This can significantly reduce the computational burden typically associated with large data structures, making viaABC aptly suitable for multidimensional data.
For instance, calculating the likelihood based on Poisson distribution over an 80 by 80 grid and across a time series in traditional Bayesian approaches would require heavy computation and becomes increasingly challenging at higher resolutions and longer timescales.
Likelihood-free ABC-based approaches handle the computational burden through the implementation of low dimensional summary statistics for high-dimensional data structures.
Nevertheless, the efficacy of these approaches is contingent upon the judicious selection of summary statistics and compatible distance metrics.
Automating this selection process has been the subject of an intense study.

Fearnhead and Prangle were among the first to effectively bypass the need for selecting summary statistics in ABC~\citep{fearnhead2010}.
They demonstrated that the posterior mean of parameters serves as an optimal summary statistic under quadratic loss  and introduced a regression-based method for its estimation.
Building on this foundation, Jiang \etal.~\citep{Jiang2017} employed dense neural network layers, while Akesson \etal.~\citep{akesson2021convolutional} used convolutional neural networks (CNN) to encode the posterior mean within the hidden layers of their architectures before incorporating it as a summary statistic in ABC frameworks.
Notably, Wang \etal.~\citep{Wang2019} leveraged a VAE model to automate summary statistic selection.
Specifically, they derived latent variables ($z^\text{sim}$) from a small sample of simulated data and then used a support vector regression model to link $\theta$ and $z^\text{sim}$, treating the latter as response variables.
Predicted $z^\text{sim}_i$ for each sampled $\theta_i$ were then used for parameter inference.

Our approach differs fundamentally from the studies described above in its learning objective.
Rather than approximating the posterior mean, viaABC is designed to create regularized and meaningful latent representations while reconstructing the original data.
This design philosophy aligns closely with the \textit{manifold hypothesis}---the principle that the underlying structure of high dimensional data resides within low dimensional manifolds embedded within the high dimensional space.
Moreover, our model architecture provides a key advantage that it can encode and analyze both spatial and temporal information within a single, unified inference framework, leveraging the inherent capabilities of the vision transformer.
In contrast, the previously described approaches are typically designed to handle either spatial or temporal data exclusively.
Although a sequential CNN that first extracts spatial features before modeling temporal dynamics could theoretically achieve comparable results.

Lastly, we demonstrated viaABC's ability to leverage contextual dependencies beyond spatial contexts. 
We showed that viaABC achieved results comparable to established Bayesian algorithms when estimating parameters of the Lotka-Volterra predator-prey model using sparse, noisy data.
In future work, we will expand our framework to enable comparison and ranking of mathematical models. 
We propose implementing hierarchical architectures that encode simulations from independent models within a single, categorized latent space. 
The resulting ensemble representations of simulated dynamics across all candidate models can, in principle, be leveraged to derive model-averaged predictions and compute model weights in conjunction with parameter inference.

\section*{Methods}
\subsection*{Data generation for the SIRS model:}
We simulated the spatial SIRS model using predefined initial conditions on an $80 \times 80$ grid, where each cell represents an individual. 
Across all simulations, we chose five infection hotspots within the grid.
The model is initialized at $t=0$ with an infected individual in the vicinity of each infection hotspot. 
The exact locations of initial infections in each simulation was varied stochastically by sampling from a uniform distribution over (-5,5) in each direction.
Susceptible individuals get infected with the following probability
$
P(\text{Infected } | \text{ Susceptible}) = 1 - e^{-\beta \cdot i \cdot \Delta t},
$
where $i$ is the number of infected neighbors and $\Delta t$ is the simulation time step.
An infected individual recovers from an infection after a fixed period $\tau_I$ and a recovered individual remain resistant to new infection for $\tau_R$ time period, which was fixed to be $1.0$.
For simulations and inference, we use general priors $Unif(0,  4.3)$ for both $\beta$ and $\tau_I$, and set the stepsize $\Delta t = 0.05$. 
Generated video is then sliced into images taken at every 20 steps form the training data.
We generate the observed data using same procedure at $\beta$ and $\tau_I = (1.0, 1.0)$.

\subsection*{Data generation and scaling for the Lotka-Volterra model:}
We simulated the Lotka-Volterra model using fixed initial conditions of $(x, y) = (1, 0.5)$. For training, the parameters $a$ and $b$ were sampled from uniform priors, $a, b \sim \text{Unif}(0, 10)$, using Latin Hypercube Sampling (LHS).
The model was numerically integrated using the \textit{RK45 ODE solver} in \textit{python}.
Variation between data scales in multivariate time series presents optimization challenges for deep learning models. Data normalization is thus a crucial pre-processing step to ensure stable training. 
A widely used normalization approach involves applying an affine transformation to the time series, i.e.. $\tilde{x} = {(x_i - m)}/{s}$ \citep{rabanser2020effectiveness, ansari2024chronos}.
We employed mean scaling by setting $m = 0$ and $s = \frac{1}{n}\sum_{i=1}^{n}|x_i|$ for both simulated and observed data generated using the Lotka-Volterra model, as it has demonstrated effectiveness in deep learning models frequently applied to practical time-series tasks \citep{salinas2020deepar, rabanser2020effectiveness, ansari2024chronos}.
During inference, parameter values were drawn from the same uniform priors, but without LHS sampling.

\textbf{Code availability:}
Code containing model definitions, training scripts, inference scripts, and notebooks used to generate our figures for reproducibility is available in a GitHub repository archived on Zenodo: 
\ULurl{https://zenodo.org/records/15776261}.

\vspace{2mm}

\begin{algorithm}
\caption{{viaABC: Variational Inference Assisted Approximate Bayesian Computation}}
\begin{algorithmic}[1]
\setstretch{1.2}
\State \textbf{Given:}
\Statex \hspace{\algorithmicindent} $\zeta(\cdot)$: a trained encoder
\Statex \hspace{\algorithmicindent} $T$: total number of iterations for termination
\Statex \hspace{\algorithmicindent} $N$: number of accepted particles per iteration
\Statex \hspace{\algorithmicindent} $k$: multiplier for the initial sampling pool ($k > 1$)
\Statex \hspace{\algorithmicindent} $q$: final quantile threshold for termination
\Statex \hspace{\algorithmicindent} $z^{\text{obs}}$: encoded observed data, i.e., $z^{\text{obs}} = \zeta(y^{\text{obs}})$
\vspace{0.5em}

\Statex \textbf{Initialization ($t=1$):}
\For{$i = 1$ to $kN$}
    \State Sample: $\theta^{(1)}_i \sim \pi(\theta)$
    \State Simulate: $y^{sim} \sim f(y \mid \theta^{(1)}_i)$
    \State Encode: $z^{sim}_i = \zeta (y^{sim})$
    \State Compute distance: $d_i = \rho(z^{\text{obs}}, z^{sim}_i)$
\EndFor
\State Let $D = \{d_i\}_{i=1}^{kN}$ and sort $D$ in increasing order
\State Re-order particles $\{\theta^{(1)}_i\}_{i=1}^{kN}$ accordingly to D
\State Set initial tolerance $\epsilon_1 = d_{(N)}$ (the $N$th smallest value in $D$)
\State Set initial particles: $\{\theta^{(1)}_i\}_{i=1}^{N}$
\Statex \textbf{Iterations ($2 \leq t \leq T$):}
\For{$t = 2$ to $T$}
    \State Compute weighted empirical variance: $\tau^2_{t-1} = 2 \cdot \text{Var}_\omega\left(\{\theta^{(t-1)}_i\}_{i=1}^N\right)$
    \For{$i = 1$ to $N$}
        \Repeat
            \State Sample $\theta^* \sim \{\theta^{(t-1)}_j\}_{j=1}^N$ with weights $\{\omega^{(t-1)}_j\}_{j=1}^N$
            \State Perturb: $\theta^{**}_i \sim \mathcal{N}(\theta^*, \tau^2_{t-1})$
            \State Simulate $y^{sim} \sim f(y \mid \theta^{**}_i)$
            \State Encode: $z^{sim} = \zeta(y^{sim})$
            \State Compute distance: $d = \rho(z^{\text{obs}}, z^{sim})$
        \Until{$d \leq \epsilon_{t-1}$}
        \State Record distance: $d^{(t)}_i = d$
        \State Record particle: $\theta^{(t)}_i\ = \theta^{**}$
        \State Update weight:
        \[
        \omega^{(t)}_i \propto \frac{\pi(\theta^{(t)}_i)}{\sum_{j=1}^{N} \omega^{(t-1)}_j \cdot \phi\left(\theta^{(t)}_i; \theta^{(t-1)}_j, \tau^2_{t-1}\right)}
        \]
        where $\phi(\cdot;\mu,\sigma^2)$ is the normal density function
    \EndFor

     \algstore{viaABC}
\end{algorithmic}
\label{alg:viaABC}
\end{algorithm}

\begin{algorithm}[h!]
\begin{algorithmic}[1]
    \algrestore{viaABC}
    \setstretch{1.2}
    \vspace{0.5em}
    	\State Normalize weights such that $\sum_{i=1}^{N} w_i^{(t)} = 1$
        \State Estimate normalizing constant:
    \[
    \hat{c}_t = \sup_{\theta} \frac{\pi_{\epsilon_t}(\theta)}{\pi_{\epsilon_{t-1}}(\theta)}
    \]

    \State Update quantile level: $q_t = \frac{1}{\hat{c}_t}$
    \State Update tolerance: $\epsilon_t = \text{Quantile}(\{d^{(t)}_i\}_{i=1}^N, q_t)$
    \If{$q_t \geq q$}
        \State \textbf{Break}
    \EndIf
\EndFor
\end{algorithmic}
\end{algorithm}

\newpage
\bibliographystyle{unsrt}
\bibliography{refs}

\newpage
\appendix

\section*{Supplementary Information}

\renewcommand{\thefigure}{S\arabic{figure}}
\setcounter{figure}{0}  

\begin{figure}[htbp]
    \centering
    \includegraphics[width=0.75\linewidth]{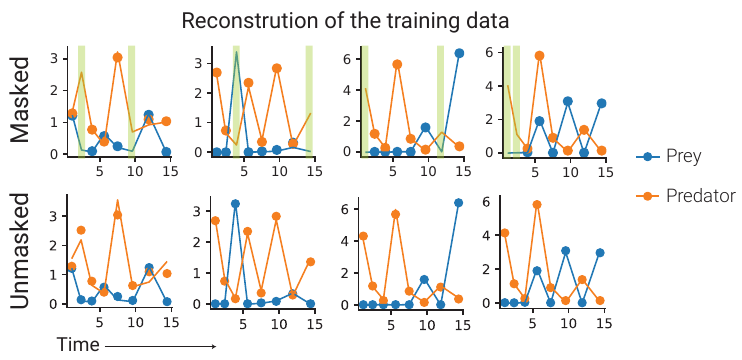}
\caption{\textbf{Masking Approach for the Lotka-Volterra model.} Comparison of reconstructions from masked and unmasked inputs.
Four randomly selected training samples are shown, with masked inputs in the first row and unmasked inputs in the second row.
For each sample, $\approx 15\%$ of the data is masked at random (\ie 2 data points) of the input is masked.
Green bars indicate the positions of the masked data points. 
Lines represent the reconstructions of the original input shown as black dots.}
    \label{fig:masking}
\end{figure}

\clearpage
 
\begin{figure}[htbp]
    \centering
    \includegraphics[width=1\linewidth]{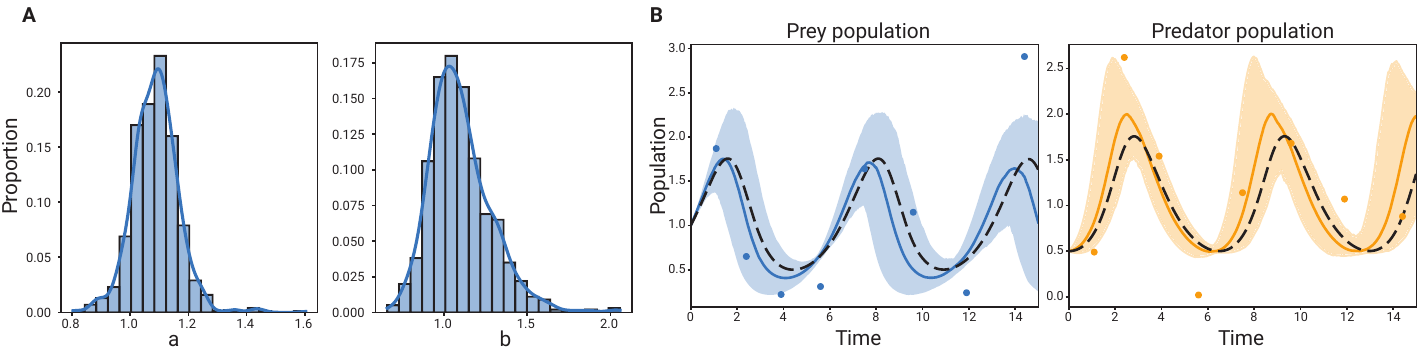}
\caption{\textbf{Inferring predator-prey dynamics using MCMC.} 
\textbf{A.} Posterior distribution of parameters $a$ and $b$ using 1,000 accepted particles from MCMC for the Lotka-Volterra system. \textbf{B.} Predicted predator-prey dynamics with 95\% credible intervals (shaded), and median trajectories (solid lines). Black dashed lines represent the ground truth dynamics.}
    \label{fig:stan_fit}
\end{figure}

\end{document}